\documentstyle[psfig,prb,aps]{revtex}
\begin{document}
\draft

\twocolumn[\hsize\textwidth\columnwidth\hsize\csname
@twocolumnfalse\endcsname

\title{
Ga$^+$, In$^+$ and Tl$^+$ impurities in alkali halide crystals: distortion trends.
}
\author{Andr\'es Aguado}
\address{Departamento de F\'\i sica Te\'orica, Facultad de Ciencias,
Universidad de Valladolid, 47011 Valladolid, Spain}
\date{}
\maketitle
\begin{abstract}
A computational study of the doping of alkali halide crystals
(AX: A = Na, K; X = Cl, Br) by ns$^2$ cations (Ga$^+$, In$^+$ and
Tl$^+$) is presented. Active clusters of increasing size (from 33 to 177 ions)
are considered in order to deal with the large scale distortions induced by the 
substitutional impurities. Those clusters are embedded in accurate quantum
environments representing the surrounding crystalline lattice. The convergence
of the distortion results with the size of the active cluster is analyced for
some selected impurity systems. The most important conclusion from this study
is that distortions along the (100) and (110) crystallographic directions are
not independent. Once a reliable cluster model is found, distortion
trends as a function of impurity, alkali cation and halide anion are identified
and discussed. These trends may be useful when analycing other cation impurities
in similar host lattices.
\end{abstract}

\vskip2pc]

\section{Introduction}
\label{aguado:intro}

Luminescent materials are used in a wide variety of technological applications.
\cite{Bla95} An important proportion of these materials is obtained by doping
a pure ionic crystal with substitutional impurities which have desirable
absorption-emission properties. The introduction of these substitutional
impurities induces a distortion of the host lattice in a local region around
the defect, due mainly to the different sizes of the impurity and host ions.
That distortion, which is different for each combination of impurity ion and 
host lattice, determines the specific lattice potential felt by the impurity,
and so the location of the impurity electronic levels inside the band gap
of the pure crystal. This brief introduction highlights the importance of
having a good understanding of the lattice distortions induced by substitutional
impurities in ionic crystals. Moreover, their experimental measurement
is a difficult task,\cite{Bar92,Pon90,Zal91} and so theoretical calculations
become an ideal complement to the experimental studies.

From the theoretical side, to obtain an accurate characterization of the local
structure around a defect is a delicate matter. First of all, the full
translational symmetry of the crystal is lost, and so Bloch's theorem can not
be directly applied. One way to avoid this problem is to duplicate a finite
region of the crystal around the impurity to recover full translational
symmetry and exploit the computational
convenience of Bloch's theorem. This is done in supercell techniques.
\cite{Pus98} The supercell size has to be chosen sufficiently large so that
the interaction between the local region influenced by the defect and its
periodical images do not interact, because that interaction would be 
nonphysical. If one forgets about Bloch's theorem, one is left with the cluster
approach, in which the doped crystal is modeled
by a finite cluster centered on the impurity and embedded in a 
field representing the rest of the host lattice. This cluster approach is the
one chosen in the present study, and has been used in the past
to study the geometrical and optical properties
of doped crystals. \cite{Che81,Win87,Kun88,Bar88,Lua89a,Lua89b,Sei91,And91,Isl92,Lua92,Miy93,Sca93,Lua93,Vis93,And93,Flo94,Bel94,Ber95,Pas95,Gry95,Llu96,Sei96,Ber96a,Ber96,Riv98,Agu98,Bar99,Agu00}
In this approach it is important to achieve an accurate description of both
the active cluster and the environment. Moreover, those two descriptions
should be consistent with each other. Usual deficiencies found in previous
works employing the cluster approach are the following: (a) an incomplete
representation of the environment. In the simplest and most frequently used 
approach, the environment is simulated by placing point charges on the
lattice sites. This procedure has to be improved in order to obtain a
realistic description of the lattice distortions around the impurity.
\cite{Bar88,Lua89a,Lua89b,Sei91,Lua92,Lua93,Flo94,Pas95,Llu96,Sei96,Agu98,Agu00}
Model potentials have been developed to represent the effects of the
environment on the active cluster, that include attractive and repulsive
quantum-mechanical terms aside from the classical Madelung term.\cite{Huz87} 
(b) An active cluster size that is too small. This is usually a problem of the
computationally expensive {\em ab initio} calculations.
Only the positions of the ions in the first shell around the impurity are
allowed to relax in most cases.\cite{Sei91,Pas95,Llu96,Sei96,Ber96,Bar99}  
However, geometrical relaxations far beyond the first shell of
neighbors can be expected for certain impurities, as suggested by
recent semiempirical simulations of solids. 
\cite{Zha93,Zha94,Isl94,Isl95,Say95,Akh95,Exn95,Cat98a,Cat98b} (c) An abrupt
connection between the active cluster and the environment. The wavefunctions and
positions of the ions in the environment are not allowed to relax, so that the
ions in the surface of the active cluster feel a wrong embedding potential. In
previous works,\cite{Agu98,Agu00} we have shown how an unphysically abrupt
connection will lead to wrong distortions independently of the intrinsic 
accuracy of the electronic structure code. (d) An incomplete self-embedding
consistency. The cluster model must be tested with calculations on the pure
crystals in order to supress systematic errors from the calculated distortions.
\cite{Agu98,Agu00}

The most expensive {\em ab initio} calculations become impractical when the
defect induces large-scale lattice distortions. On the other hand, semiempirical
calculations employing pair potentials are not completely reliable due to the
problem of transferability to environments different from those in which they
were generated. In this paper we employ the {\em ab initio} Perturbed Ion (aiPI)
\cite{Lua90a,Lua90b,Pue92,Lua92a,Lua93a}model to study the lattice distortions 
induced by Ga$^+$, In$^+$ and Tl$^+$ impurities in NaCl, NaBr, KCl and KBr.
With the use of this method we circunvent all the incoveniences listed above:
(a) The active cluster is embedded in a quantum environment
represented by the {\em ab initio} model potentials of Huzinaga 
{\em et al.} \cite{Huz87}; (b) The computational simplicity of the PI model
allows for the geometrical relaxation of several coordination shells around the
impurity;\cite{Lua92,Lua93,Agu98,Agu00} (c) The local region around the
defect in which structural rearrangements are important is connected to the
frozen crystalline environment by a smooth interface of ions whose wavefunctions
are allowed to selfconsistently relax; (d) Parallel cluster model calculations
on the pure crystals are performed in order to obtain trustworthy distortions.
By studying several related systems we try to identify
systematic distortion trends that might be useful in
later theoretical studies of doped crystals similar to those here considered.
In a previous publication\cite{Agu00} we considered the case of ns$^2$ anionic
substitutional impurities. Now we complete the study of ns$^2$ substitutional
impurities in alkali halide crystals with a consideration of cationic
impurities.

The remainder of this paper is organized as follows: 
In Section \ref{aguado:theory}
we describe the active clusters which have been used to model the doped systems.
In Section \ref{aguado:results} we 
present and discuss the results of the calculations, and Section 
\ref{aguado:summary} summarizes
the main conclusions.

\section{Cluster model} 
\label{aguado:theory}

The {\em ab initio} Perturbed Ion model is a particular application of the
theory of electronic separability of Huzinaga and coworkers
\cite{Huz71,Huz73} to ionic solids
in which the basic building blocks are reduced to single ions.
The PI model was first developed for perfect crystals.\cite{Lua90a}
Its application to the study of impurity centers
in ionic crystals has been described in refs. \onlinecite{Lua92,Lua93,Agu98}, 
and we refer to those papers for a full account of the method. 
In brief, an active cluster containing the impurity is considered, and
the Hartree-Fock-Roothaan (HFR) equations\cite{Roo63} for 
each ion in the active cluster are solved in the field of the other ions.
The Fock operator includes, apart from the usual intra-atomic terms, an
accurate quantum-mechanical crystal potential
and a lattice projection operator which accounts for the energy contribution
due to the overlap between the wave functions of the ions.\cite{Fra92}
The atomic-like HFR solutions are used to describe the ions in the 
active cluster in an iterative stepwise procedure.
The wave functions of the lattice ions external to
the active cluster are taken from a PI calculation for the perfect crystal
and are kept frozen during the embedded-cluster calculation.
Those wave functions are explicitely considered for ions 
up to a distance $d$ from the center of the
active cluster such that the quantal contribution from the most distant frozen
shell to the effective cluster energy is less than 10$^{-6}$ hartree.
Ions at distances 
beyond $d$ contribute to the effective energy of the active cluster just
through the long-range Madelung interaction, so they are represented by point
charges. At the end of the calculation, the ionic wave functions
are selfconsistent within the active cluster and consistent with the frozen
description of the rest of the lattice. The intraatomic Coulomb correlation,
which is neglected at the Hartree-Fock level, is computed as a correction by
using the Coulomb-Hartree-Fock (CHF) model of Clementi.\cite{Cle65,Cha89} 
The polarization contribution is computed by using the polarizable ion model 
devised by Madden and coworkers,\cite{Mad96,Jem99}
as explained in previous publications.\cite{Agu99,Agu00b} In-crystal
polarizabilities were obtained from Ref. \onlinecite{Fow85}.

Now we describe the cluster models of increasing complexity employed to simulate
the impurity systems. All of them are embedded in accurate quantum environments
as indicated in the paragraph above. The active clusters are split up into two
subsets, which we call {\bf C}$_1$ and {\bf C}$_2$ following Refs.
\onlinecite{Lua92,Lua93}. Both the positions and wavefunctions of the ions in the
inner {\bf C}$_1$ subset are allowed to relax. The positions of the ions in the
outer {\bf C}$_2$ subset are held fixed during the optimization process, but 
their wavefunctions are selfconsistently optimized. Thus, the {\bf C}$_2$ subset
provides a smooth interface connecting the {\bf C}$_1$ region, where distortions
are important, to the frozen crystalline environment. In practice, {\bf C}$_2$
contains all those ions which are first neighbors of the ions in {\bf C}$_1$ 
and are not already contained in {\bf C}$_1$. In Table I we show the lattice
ions included in the {\bf C}$_1$ subset for the different cluster models. It has
to be understood that the impurity central ion is included in all cluster 
models, and that each cluster model is formed by adding the ions shown in Table
I to the preceding cluster model. In each case,
the geometrical relaxation around the impurity
has been performed by allowing for the independent
breathing displacements of each shell of ions, and minimizing the total
energy with respect to those displacements until the effective
energies are converged up to 1 meV. A downhill simplex algorithm\cite{Wil91}
was used. For the description of the ions we have used large STO basis sets, 
all taken from Clementi-Roetti\cite{Cle74} and McLean-McLean\cite{McL81} tables.
The calculations have been performed
by employing the experimental lattice constants\cite{Ash76} to describe the
geometrically frozen part of the crystals. 

The self-embedding consistency of the method has been checked for all the
different cluster models and found to be of a quality similar to that found in
previous papers.\cite{Agu98,Agu00} By this we mean that if the pure crystal is
represented by one of the cluster models enumerated above
(that is, if the central impurity is replaced
by the alkali cation corresponding to the pure crystal), the results of the
cluster model calculations closely agree with those from a PI calculation for
the pure crystal, where all cations (or anions) are equivalent by translational
symmetry. Nevertheless, the self-embedding consistency is never complete.
In order to supress systematic errors from the distortions calculated with
the cluster method, the radial displacements of each shell have been 
calculated using the following formula:
\begin{equation}
\Delta R_i = R_i(Imp^+:AX) - R_i(A^+:AX),
\end{equation}
where R$_i$ (i=1, 2, 3, 4, 7, 8) refer to the radii of the first, second, third,
fourth, seventh and eigth shells around the impurity in the AX crystal, 
A = Na, K, X = Cl, Br, and Imp$^+$ = Ga$^+$, In$^+$, Tl$^+$.
Thus both systems (pure and doped crystals) are treated in eq. (1) on equal
foot so that any inaccuracy in the calculated distortions is completely due to
the electronic structure method.

\section{Results and discussion}
\label{aguado:results}

\subsection{Influence of the active cluster size}

We show in Table II the lattice distortions calculated with the different 
cluster models described in the previous section, for the NaCl:Tl$^+$ system.
As the distortion trends for different impurity systems will be discussed in 
the next subsection, here we just point out two general results:
(a) the distortion in the first coordination
shell, $\Delta R_1$, depends upon an explicit consideration of the distortion
in the fourth coordination shell, $\Delta R_4$, that is the distortion
along the (100) crystallographic direction propagates at least until next
nearest neighbor positions. This is clearly seen in Table II by comparing the
value of $\Delta R_1$ in cluster model C with
the value of $\Delta R_1$ in cluster models D and E. When the positions of the
ions in the fourth coordination shell are held fixed, the $\Delta R_1$
expansion can not attain its optimal value;
(b) the distortions along (100) and (110) 
crystallographic directions are coupled. This means that the calculated 
distortions in the first and fourth coordination shells depend upon an
explicit inclusion of the second and eigth coordination
shells in the active clusters. This
is seen in the value of $\Delta R_1$ when passing from model A to model B and
in $\Delta R_1$ and $\Delta R_4$ when passing from model E to model F. These
conclusions are independent of the specific system studied. For example, we
show in Fig. 1 the evolution of $\Delta R_1$, as a function of the active
cluster complexity, for the three cationic impurities in NaCl. We may
appreciate that the three lines are neatly parellel to each other, supporting
the validity of our conclusions. Thus, the errors in $\Delta R_1$, due only to 
a small active cluster size (that is not directly associated with the accuracy
of the electronic structure code), can be roughly estimated to be 1\%.

A magnitude which is very difficult to estimate theoretically
is the defect formation energy.
At low pressure and temperature conditions, the formation of the
impurity centers should be discussed in terms of the internal energy difference
for the exchange reaction\cite{Flo94}
\begin{equation}
(A^+:AX)_s + Imp^+_g \rightleftharpoons (Imp^+:AX)_s + A^+_g,
\end{equation}
where the $s$ and $g$ subindexes refer to solid and gas phases, respectively.
The Madelung energy term, which is the most important contribution to cohesion
in ionic crystals, is minimized when the ions are in their perfect 
crystallographic positions, so the lattice distortions described above will
tend to destabilize this term. The defect will be stable only if the
electronic contributions can compensate for that Madelung energy loss.
The detailed energy balance contains then nonnegligible contributions 
from energy terms, like the dispersion interactions,
that are not so important for the determination of structural properties.
The dispersion interactions, as well as any relativistic energy
terms, are not included in the aiPI formalism,\cite{Agu00} so we do not expect
the formation energies calculated from it to be wholly reliable. Nevertheless,
the contribution to the defect formation energy coming from elastic 
relaxation of the lattice is meaningful, and we show its evolution as a function
of cluster complexity in Fig. 2, for the case of Ga$^+$, In$^+$ and Tl$^+$
impurities in NaCl. Specifically, we show the defect formation {\em energy
differences} (stabilization energies), 
calculated by substracting the formation energies calculated
with a given cluster model and its predecesor. As a reference state for the
cluster model A we took a smaller active cluster containing just 7 ions, namely
the impurity and its 6 nearest neighbor Cl$^-$ anions, which were held fixed
at their lattice positions. This cluster can be considered the limit case of
those shown in the previous section, where the {\bf C}$_1$ subset is formed
by the impurity cation and the {\bf C}$_2$ subset (the interphase) is formed
just by the six first neighbors of the impurity. The figure seems to suggest 
that the first shell distortions sligtly destabilize the defects, while
second shell distortions largely stabilize them. This apparently paradoxical
result is just a consequence of our stepped construction of cluster models.
If we include the second shell distortions alone, the net effect is also a
destabilization. This simply shows that the distortions in the first and
second shells, where short-range overlap interactions with the impurity are
important, should be included together in the calculation. Structural 
relaxation of those two shells provide the largest contribution to elastic
relaxation. We also appreciate nonnegligible stabilization energies when the
third, fourth, and eigth coordination shells are allowed to relax, giving an 
idea of the
difficulty of obtaining well converged theoretical formation energies.

\subsection{Distortion trends}

In this section we show the calculated distortions induced by substitutional
Ga$^+$, In$^+$ and Tl$^+$ impurities in NaCl, NaBr, KCl and KBr crystals. For
all these calculations we employ the cluster model F, which is the most
complete one from those considered in previous sections.
These distortions, collected in Table III, are the
main quantitative result from our study. From that table we can extract
interesting distortion trends, that we describe in the following:

{\em First shell distortion.} 
This shell undergoes an expansion in all cases, with the only exception of the
Ga$^+$ impurity in K salts. In a previous publication\cite{Agu00} we showed how
these distortions can be rationalized in terms of simple geometric arguments
involving ion sizes. As a rough measure of ion size we employed the quantity
r = $(<r^2>)^{1/2}$, where the expectation value is taken over the outermost 
orbital of the ion, and is calculated from the crystal-consistent ionic wave
functions obtained through the aiPI calculations. Then the first shell
distortions were shown to be related to the differences
$\delta$ =r(Imp$^+$:AX)-r(A$^+$:AX),
where r(Imp$^+$:AX) is the radius of the Ga$^+$, In$^+$ or Tl$^+$ impurities 
in the AX crystal, and r(A$^+$:AX) is the radius of the alkali cation in the
pure AX crystal. A similar analysis here shows that $\delta$ is negative only
for KCl:Ga$^+$ and KBr:Ga$^+$, giving an explanation to the inwards relaxation
of the first shell in those two cases. More in general, for a fixed crystal
$\delta$ increases in the series Ga$^+ \rightarrow$ In$^+ \rightarrow$ Tl$^+$,
as should be expected. The $\Delta R_1$ distortions for any given crystal follow
the same trend. For a given impurity and fixed halide anion, the first shell 
expansion is larger the smaller the size of the alkali cation. This trend
suggests what kind of distortions can be expected in Li or Rb host lattices. For
example, one can expect a small inwards relaxation for RbCl:In$^+$, similar to
that found for KCl:Ga$^+$, and more important distortions for Li salts, that
perhaps will ask for larger active cluster sizes. For those systems where an
expansion is observed, if we fixed the impurity cation and the alkali cation,
the expansion is larger the larger the size of the anion. Thus one can expect
smaller expansions for NaF and KF salts and larger ones for NaI and KI salts.
For those cases where a contraction is observed, the contraction is
larger the smaller the size of the anion.

{\em Second shell distortion.} 
The displacement of the second shell is always an expansion. For any given
host lattice, that expansion is larger the larger the impurity size, because
of the increase in cation-cation overlap repulsion. If the impurity and the
halide anion are fixed, the expansion increases with decreasing alkali size.
This trend suggests again larger $\Delta R_2$ expansions in Li salts and
smaller ones in Rb salts. If the impurity and the alkali cation are fixed, the
expansion is larger the smaller the anion size, because cation-cation
overlap decreases with increasing anion size. Then, larger $\Delta R_2$
expansions are expected in NaF and KF salts and smaller ones for NaI and KI
salts.

{\em Third shell distortion.} 
This shell experiences a small contraction in all cases.
For a given crystal, the contraction is larger the larger the size of the
impurity. If we fix the impurity and the halide anion, the contraction 
increases with decreasing alkali size. Then, larger contractions should be
expected in Li salts, and smaller ones in Rb salts. If the impurity and the
alkali cation are fixed, the contraction is larger the smaller the size of the
halide anion, so larger contractions are expected for fluoride salts. As the
direct overlap interactions between the impurity and the ions in the third
coordination shell are negligibly small, this contraction has to be considered
an indirect doping effect, mediated by the distortions induced on the first and
second shells. The small contraction of the third shell
optimizes the Madelung energy around the impurity and also serves to pack more
efficiently the ions in response to the outward motion of the ions in the first
and second shells.

{\em Fourth and seventh shell distortions.} 
The fourth shell experiences an expansion, except in the case of KCl:Ga$^+$. 
The $\Delta R_4$ displacements proceed 
along the same (100) crystallographic direction as the $\Delta R_1$ 
displacements, and thus are clearly induced by the first shell distortions, as
indicated in the previous subsection. The distortion trends of this shell are
then found to be the same as those of the first shell. The displacements of
those ions in the seventh coordination shell are always negligibly small,
indicating that the distortions propagate along the (100) direction just until
next-nearest neighbor positions.

{\em Eigth shell distortion.}
Similarly to the previous paragraph, the eigth shell undergoes an expansion
induced by the expansion of the second shell, as both displacements proceed
along the (110) crystallographic directions. Thus the distortion trends are the
same as those found for the second shell.

\section{Summary}
\label{aguado:summary}

A study of the local lattice distortions
induced by substitutional Ga$^+$, In$^+$ and Tl$^+$ impurities in 
NaCl, NaBr, KCl and KBr crystals has been reported. For that purpose we have
considered active clusters of increasing complexity, with a number of ions 
varying between 33 and 177 ions, embedded in accurate quantum environments 
representing the rest of the crystal. The local distortions
obtained extend beyond the first shell of neighbors in all cases.
Thus, the assumptions frequently employed in impurity calculations,
which consider the active space as formed by the central impurity plus its
first coordination shell only, should be taken with some care. Moreover, we
have found that the distortions along the (100) and (110) crystallographic
directions are not independent from each other, and also that the first shell
distortion is not converged if the positions of the ions in the fourth 
coordination shell around the impurity are not allowed to relax. Once we have
obtained a reliable cluster model, distortion trends have been identified and 
discussed. In those cases where the size of the impurity is larger than that of
the alkali cation, the impurity induces an expansion propagating along (100) 
and (110) crystallographic directions until next-nearest neighbor positions. 
There is also an indirect contraction along the (111) crystallographic
direction, that probably affects only to the nearest neighbor ions in that 
direction, and which serves to partly compensate the
Madelung field around the impurity and to pack more
efficiently the ions in response to the outward motion of the first
and second shells. In those cases where the size of the impurity is smaller than
that of the alkali cation of the host lattice, the only significant change in 
these trends is the inwards relaxation along the (100) direction. The dependence
of the several shell distortions upon a change of alkali cation, halide anion,
and impurity cation have been described. From this description one can advance
expected distortions for closely related systems.

$\;$

$\;$

{\bf Acknowledgements:} 
Work supported by DGES (PB98-0345) and Junta de
Castilla y Le\'on (VA28/99). 
The author is greatly indebted to Angel Mart\'\i n
Pend\'as and Miguel Alvarez Blanco for providing him with an improved version
of the aiPI fortran code. 

{\bf Captions of Tables}

$\;$

{\bf Table I.}
Description of the lattice ions included in the {\bf C}$_1$ subset for the
different cluster models. The degeneracy (number of symmetry equivalent ions)
of each coordination shell is shown, together with the charge of the ions
forming that shell, and the total number of atoms in the active cluster (ions
in {\bf C}$_1$ and {\bf C}$_2$ subsets).

{\bf Table II.} 
Lattice distortions (in \AA) and percentage distortion values obtained with
the different cluster models described in the text.

{\bf Table III.}
Lattice distortions (in \AA) of several coordination shells around Ga$^+$,
In$^+$ and Tl$^+$ impurities in four alkali halide crystals, employing cluster
model F as described in the text.

$\;$

{\bf Captions of Figures}

$\;$

{\bf Figure 1.}
First-shell distortions (in \AA) induced by Ga$^+$, In$^+$ and Tl$^+$
impurities in NaCl, as a function of the complexity of the cluster model. Note
the similar behavior of the three curves.

{\bf Figure 2.}
Defect formation energy stabilization (in eV) as a function of cluster
complexity.

\newpage

\onecolumn[\hsize\textwidth\columnwidth\hsize\csname
@onecolumnfalse\endcsname

\begin{table}[t]
\begin{center}
\begin{tabular} {|c|c|c|c|c|}
\hline
Cluster Model & Lattice site & Degeneracy & Charge & N \\
\hline
A & ($\frac{1}{2}$,0,0) & 6 & - & 33 \\
B & ($\frac{1}{2}$,$\frac{1}{2}$,0) & 12 & + & 57 \\
C & ($\frac{1}{2}$,$\frac{1}{2}$,$\frac{1}{2}$) & 8 & - & 81 \\
D & (1,0,0) & 6 & + & 87 \\
E & ($\frac{3}{2}$,0,0) & 6 & - & 117 \\
F & (1,1,0) & 12 & + & 177 \\
\end{tabular}
\end{center}
\end{table}

\begin{table}[t]
\begin{center}
\begin{tabular} {|c|c|c|c|c|c|c|}
\hline
Cluster model & $\Delta R_1$ & $\Delta R_2$ & $\Delta R_3$ & $\Delta R_4$ & $\Delta R_7$ & $\Delta R_8$ \\
\hline
A & 0.165 & -- & -- & -- & -- & -- \\
  & 5.96\% & -- & -- & -- & -- & -- \\
B & 0.152 & 0.092 & -- & -- & -- & -- \\
  & 5.40\% & 2.27\% & -- & -- & -- & -- \\
C & 0.154 & 0.083 & -0.063 & -- & -- & -- \\
  & 5.49\% & 2.08\% & -1.31\% & -- & -- & -- \\
D & 0.174 & 0.081 & -0.058 & 0.053 & -- & -- \\
  & 6.23\% & 2.03\% & -1.20\% & 0.93\% & -- & -- \\
E & 0.175 & 0.080 & -0.060 & 0.054 & 0.006 & -- \\
  & 6.27\% & 2.00\% & -1.25\% & 0.95\% & 0.07\% & -- \\
F & 0.155 & 0.087 & -0.064 & 0.048 & 0.004 & 0.031 \\
  & 5.52\% & 2.18\% & -1.33\% & 0.84\% & 0.04\% & 0.38\% \\
\end{tabular}
\end{center}
\end{table}

\begin{table}[t]
\begin{center}
\begin{tabular} {c|c|ccc|c|ccc|}
\hline
& NaCl & Ga$^+$ & In$^+$ & Tl$^+$ & NaBr & Ga$^+$ & In$^+$ & Tl$^+$ \\
\hline
& $\Delta R_1$ & 0.086 &  0.136 &  0.155 & $\Delta R_1$ & 0.094 & 0.145 & 0.167 \\
& $\Delta R_2$ & 0.058 &  0.085 &  0.090 & $\Delta R_2$ & 0.045 & 0.068 & 0.072 \\
& $\Delta R_3$ & -0.041 & -0.061 & -0.064 & $\Delta R_3$ & -0.033 & -0.045 & -0.051 \\
& $\Delta R_4$ & 0.025 &  0.043 &  0.048 & $\Delta R_4$ & 0.028 & 0.047 & 0.057 \\
& $\Delta R_7$ & -0.001 & -- &     0.004 & $\Delta R_7$ & -0.006 & 0.005 & 0.007 \\
& $\Delta R_8$ & 0.021 &  0.031 &  0.033 & $\Delta R_8$ & 0.015 & 0.023 & 0.024 \\
\hline
& KCl & Ga$^+$ & In$^+$ & Tl$^+$ & KBr & Ga$^+$ & In$^+$ & Tl$^+$ \\
\hline
& $\Delta R_1$ & -0.019 &  0.055 & 0.066 & $\Delta R_1$ & -0.012 & 0.064 & 0.078 \\
& $\Delta R_2$ & 0.009 &   0.032 & 0.039 & $\Delta R_2$ & 0.004 & 0.025 & 0.033 \\
& $\Delta R_3$ & -0.016 & -0.033 & -0.035 & $\Delta R_3$ & -0.009 & -0.024 & -0.029 \\
& $\Delta R_4$ & -0.007 &  0.020 & 0.022 & $\Delta R_4$ & --    & 0.023 & 0.024 \\
& $\Delta R_7$ &  -0.007 & -- & -- & $\Delta R_7$ &    -0.006  & 0.003 & 0.006 \\
& $\Delta R_8$ &  0.007 &  0.015 & 0.016 & $\Delta R_8$ &  0.003 & 0.009 & 0.010 \\
\end{tabular}
\end{center}
\end{table}

\newpage

\begin{figure}
\psfig{figure=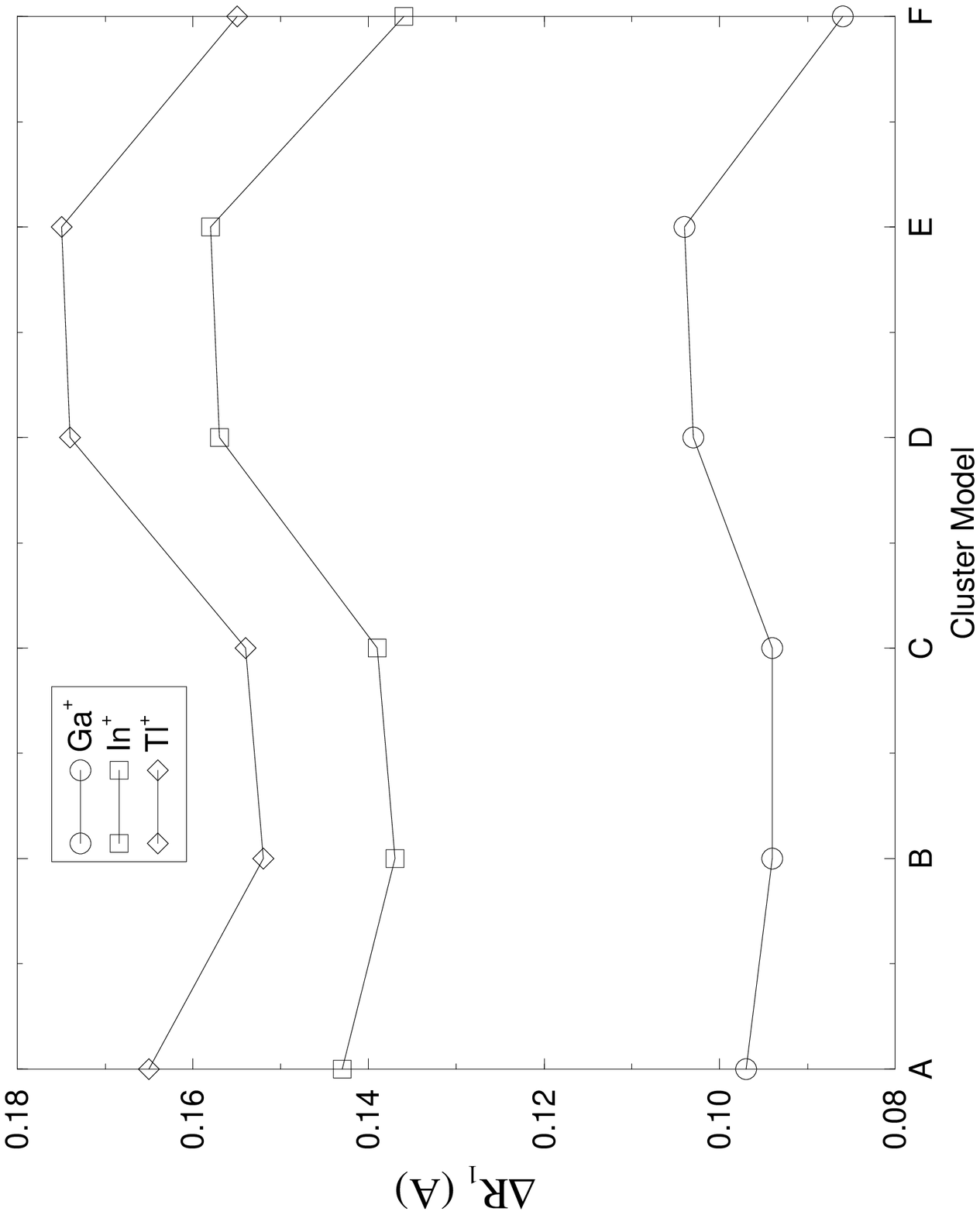}
%%\label{fig:naina:isomers}
\end{figure}

\newpage

\begin{figure}
\psfig{figure=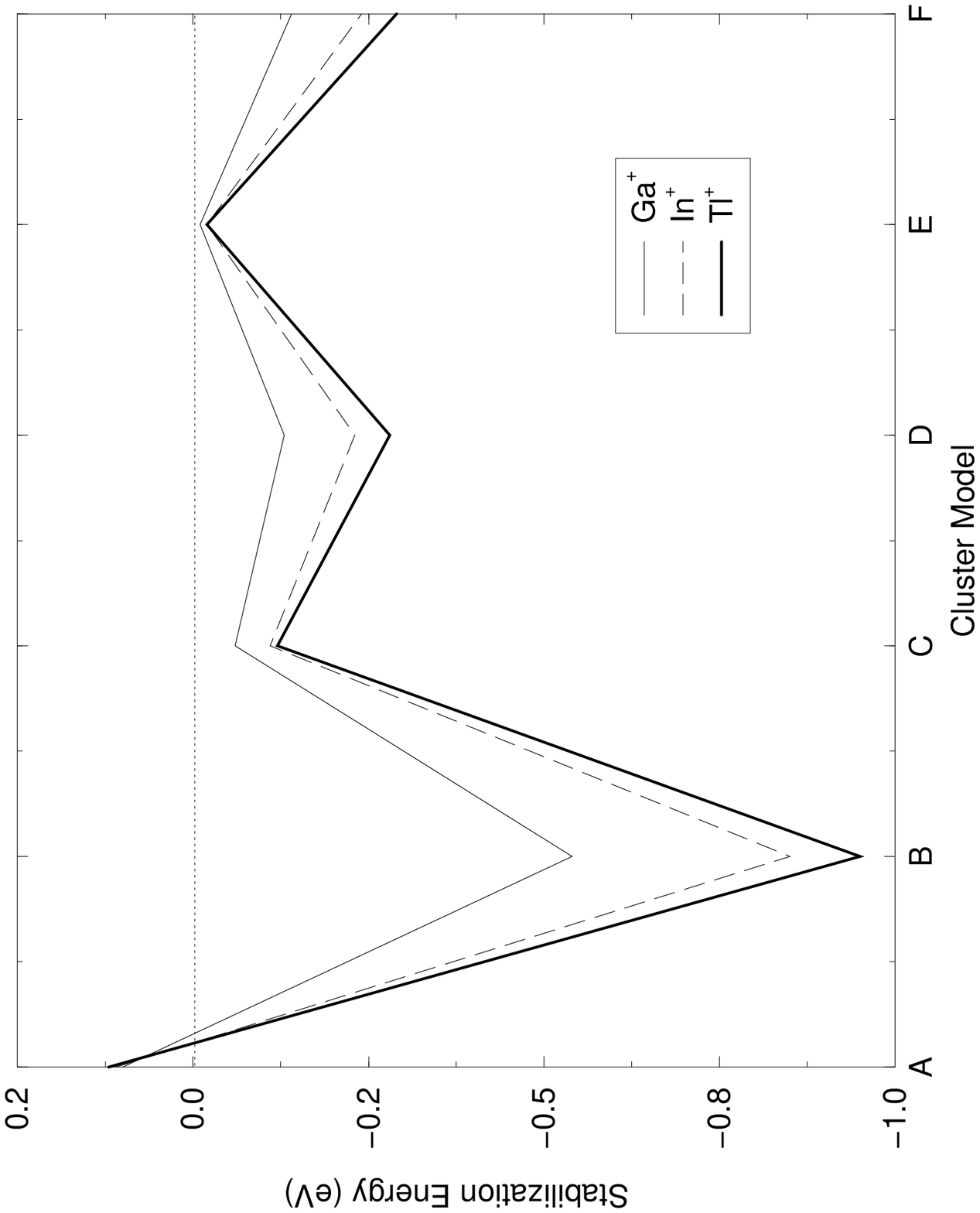}
%%\label{fig:naina:isomers}
\end{figure}

\end{document}